# Characterisation of Ferromagnetic Rings for Zernike Phase Plates using the Aharonov-Bohm effect

C. J. Edgcombe[a]*, A. Ionescu[a], J. C. Loudon[b], A. M. Blackburn[c], H. Kurebayashi[d] and C.H.W. Barnes[a]

[a] TFM Group, Department of Physics, JJ Thomson Ave, Cambridge CB3 0HE, Cambridge, UK
[b] Electron Microscopy Group, Department of Materials Science, Pembroke Street, Cambridge CB2 3QZ, UK
[c] Hitachi Cambridge Laboratory, JJ Thomson Avenue, Cambridge CB3 0HE, UK
[d] Microelectronics Group, Department of Physics, JJ Thomson Ave, Cambridge CB3 0HE, UK
*Corresponding author; email cje1@cam.ac.uk, phone +44 1223 337335, fax +44 1223 363263

**Abstract**
Holographic measurements on magnetised thin-film cobalt rings have demonstrated both onion and vortex states of magnetisation. For a ring in the vortex state, the difference between phases of electron paths that pass through the ring and those that travel outside it was found to agree very well with Aharonov-Bohm theory within measurement error. Thus the magnetic flux in thin-film rings of ferromagnetic material can provide the phase shift required for phase plates in transmission electron microscopy. When a ring of this type is used as a phase plate, scattered electrons will be intercepted over a radial range similar to the ring width. A cobalt ring of thickness 20 nm can produce a phase difference of pi/2 from a width of just under 30 nm, suggesting that the range of radial interception for this type of phase plate can be correspondingly small.

1. **Introduction**

In transmission electron microscopy (TEM) for many biological and medical applications, and especially for cryomicroscopy, it is desirable to minimise the electron dose to the specimen needed for a given image contrast. Many biological specimens have the property that when placed in an electron stream, they modify the exit wave function mainly by increasing the phase change through the specimen (relative to free-space propagation), with little change in amplitude. Such specimens are known as 'phase objects', from the corresponding behaviour of optical objects [1]. When the image of such a specimen is recorded by a detector that responds solely to intensity, the in-focus image shows little contrast. However, if the phase modulation can be converted to amplitude modulation that can be imaged by the detector, then a substantial increase in contrast at focus can be expected, and a corresponding reduction in the required electron dose should be possible.

Devices intended to achieve this conversion of modulation are known as phase plates [1] and are designed to alter the difference between the phases integrated along electron paths (a) near the electron-optical axis (the direct beam) and (b) following scattered electrons. The many types of plate that have been built or proposed include, in a chronological sequence with representative references: Foucault plates (or knife edges) that extend over a complete half-plane [2,3,4] or part of one [5], magnetised rings producing a phase difference by the Aharonov-Bohm effect [6,7], discs with controlled thickness and central apertures [8,9,10], electrostatic systems with ring or cylinder electrodes [11,12,13], magnetised wires [14], and strong electromagnetic fields [15]. Structures that are rotationally symmetric (Zernike geometry) have the benefit that electrons are treated similarly for all azimuthal directions of scattering. The use of such phase plates in combination with aberration-corrected microscopes is expected to generate optimal phase contrast [16] over the entire resolution range of the microscope, while by minimizing the diameter of the aperture of a Zernike phase plate the effective lower limit of spatial frequency response can be substantially reduced.

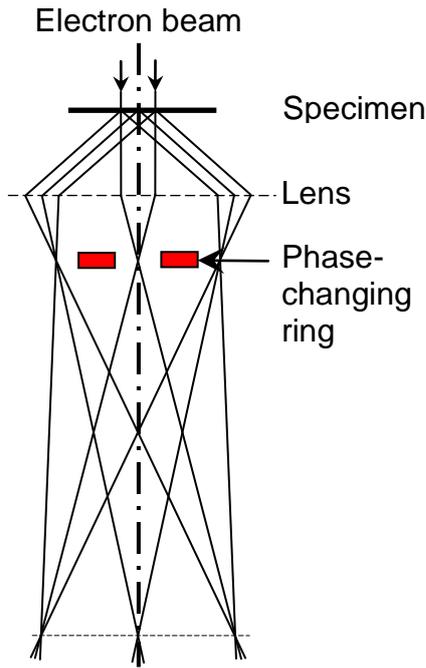

Fig. 1 Direct and scattered electron trajectories in a TEM, with a magnetic ring located in the back focal plane of the objective lens.

One possibility for a phase plate is obtained by using the phase difference produced by a ring of ferromagnetic material, magnetised with a continuous loop of flux (Fig.1). It was realised earlier [6] that the Aharonov-Bohm effect would cause a phase difference between electron paths through such a ring and those outside it. In discussions without knowledge of the earlier work, J A C Bland suggested [17] that thin-film rings made of cobalt might resist axial fields of the order of magnitude required. Within the material, the flux density has the saturation value, but the magnitude of flux needed is tiny (~$10^{-15}$ Wb) so the area of cross-section needed is correspondingly small. The obvious question is whether such a ring of flux can be maintained in the presence of the axial (out-of-plane) field of an objective lens. If this can be achieved, then a ring of this type will offer not only the benefit of rotational symmetry, but also that of very small radial interception. It is easily estimated that the radial width needed for a cobalt ring can be as small as 30 nanometres.

Here we describe relevant theory, simulations, fabrication and holographic measurements for magnetised thin-film rings of cobalt whose dimensions are comparable with those of the aperture in Zernike phase plates for TEM. These tests confirm the known states of magnetisation in narrow thin-film rings and show additional quantitative details of magnetisation in a solid disc.

## 2. Theory for use of magnetic rings as phase plates

When electrons travel from a point p to a point q by a path L through a region where the vector potential is **A**(L), there is a contribution to the change of phase of their wave function of $(e/\hbar)\int_p^q A \cdot dL$, where *e* is the magnitude of the charge on the electron and $\hbar$ is the reduced form of Planck's constant [18]. Then if electrons start from the same point on the cathode, travel by two paths $L_1$ and $L_2$ and arrive at the same point on the detector, the changes of phase for the two paths differ in general by an angle $\Delta\theta$, given by

$$(\hbar/e)\Delta\theta = \int_p^q A \cdot dL_2 - \int_p^q A \cdot dL_1 = \oint^{p2q1p} A \cdot dL = \int^S \nabla \times A \cdot dS = \int^S B \cdot dS = \Phi(S) \quad (1)$$

where p2q1p denotes the closed path from p by path L2 to q then by path L1 to p, **S** is a surface bounded by $L_1$ and $L_2$, **B** is the magnetic flux density and $\Phi$ is the magnetic flux through this surface. This non-zero $\Delta\theta$ follows from the requirement that electromagnetic quantities should be independent of the choice of gauge and is known as the Aharonov-Bohm effect [19,20]. For the present application we consider paths differing mainly by displacement radially from the axis of a microscope, such as those of direct and scattered electrons (Fig. 1). The direction of flux that produces this effect in one pair of paths is perpendicular to both the electrons' paths and the radial direction. Consideration of pairs of paths with this radial displacement at all angles around the axis shows that the required flux distribution consists of a continuous loop of flux, of the same magnitude at all azimuthal angles.

## 3. Properties of ferromagnetic thin-film rings

In a ferromagnetic material, the exchange interaction acts over distances of the order of nanometres to align neighbouring electronic spins. This effect favours transverse alignment of spins and constrains the structures of domains that can exist stably. When a narrow ferromagnetic ring is subjected to a sufficiently strong magnetic field in-plane, the directions of spins form two semi-circular arcs (Fig. 2a), meeting in head-to-head and tail-to-tail domain walls, as would be expected in a ring of macroscopic size. This state has been named the 'onion' state, from its resemblance to an onion bisected along its axis. However, in rings with suitably small dimensions, it is found that on reducing and reversing the in-plane field, another state can be obtained, containing a continuous loop of flux [21]. This is known as the 'vortex' state (Fig. 2b). Given suitable material and dimensions, it can be stable against applied in-plane fields up to ± 300 mT or more. A simulation of the magnetization $M_x$ resulting from applied in-plane flux density $B_x$ for a ring of outer diameter 2 micrometres and width and thickness 100 x 20 nanometres is shown in Fig. 3. Over the central line (coloured red online), on which $M_x$ is approximately proportional to $B_x$ up to an applied field of about ±130 mT, the ring is in the vortex state.

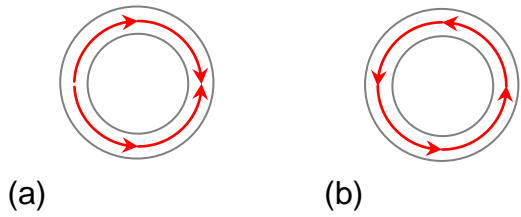

Fig. 2 Distributions of flux for rings in (a) the 'onion' state and (b) the vortex state.

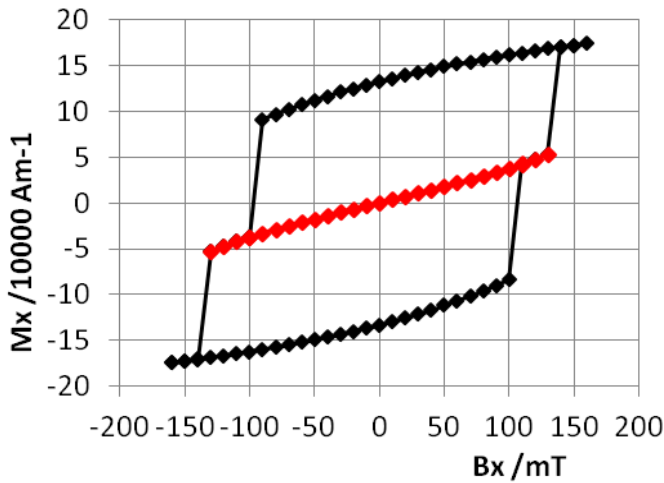

Fig. 3 Simulated in-plane magnetization $M_x$, as function of in-plane flux density $B_x$ applied to a circular cobalt ring of outer diameter 2 micrometres, width 100 nanometres and thickness 20 nanometres.

As proposed earlier [22,7] we suggest that a ring magnetised in the vortex state be inserted in the back focal plane of an electron lens (Fig. 1), with the ring dimensions chosen so that an unscattered electron passes through the ring, while an electron from the same point on the cathode and scattered from the specimen at a sufficiently large angle passes outside it. If these two electrons are brought to meet the detector at the same point, their paths define a complete loop and differ in phase change by $(e\Phi/\hbar)$ as described above. Here $\Phi$ is the circumferential (or azimuthal) flux through a section of the ring. There is no applied circumferential flux density outside the section of the ring, so the phase difference does not depend on the precise location of electron trajectories near the ring and is the same between any path within the ring and any outside it. The phase difference also has no dependence on the energy of the electron beam. For use as a phase plate, the diameter of the ring must satisfy the same condition as for the small aperture in a Zernike plate: it should allow direct electrons to pass through the ring and scattered ones to pass outside it. There is no need for further material in the beam tunnel outside this diameter, other than any needed to support the ring and to neutralise charge.

The area of cross-section required to produce a particular phase change can be calculated easily. Since $\Delta\theta = e\Phi/\hbar$ as described above, the flux required for a phase change of $\pi/2$ is $1.03 \times 10^{-15}$ Wb. The saturation magnetisation of cobalt in bulk is 1.79 T [23], so, if the same value is achieved in a thin film, a ring cross-section of 580 nm$^2$ is enough to produce the required phase change. Then if, for instance, the thickness of the ring is chosen as 20 nm, the radial width of the ring need not be more than 30 nm, and interception of the scattered electrons is limited (in principle) to this small range of radii.

### 3.1 Simulation

The magnetic behaviour of thin-film rings has been computed with the micromagnetics simulation package OOMMF [24]. This package has many capabilities, but here we report on switching between onion and vortex states by a steady field applied purely in-plane. The material parameters were chosen to represent polycrystalline cobalt deposited with hexagonal crystallites and were: Exchange stiffness constant A = 3 x 10$^{-11}$ Jm$^{-1}$, saturation magnetization $M_s$ = 1.4 x 10$^6$ Am$^{-1}$, uniaxial anisotropy constant $K_1$ = 5.2 x 10$^5$ Jm$^{-3}$. The cell size used was 5 x 5 x 20 nm.

The molecular-beam epitaxy (MBE) process deposits polycrystalline cobalt on SiN in the form of grains, and it is known that when a similar deposition process is used for iron the grains have typical dimensions of the order of 14 nanometres [25]. In a complete ring there are many such grains, whose crystalline orientation varies randomly. The resulting structure has an inherent fluctuation of local directions of easy magnetization, with a total initial magnetization that is unpredictable in magnitude and direction. A consequence of this random initial magnetization is that in experiments on a group of rings of the same dimensions, a specific switching transition may be found to occur over a range of values of field. It may be possible to model this behaviour in the simulation by providing a grain map, but at the time of calculation the typical grain size for Co was not known. To model the behaviour very simply, the file Oxs_UniaxialAnisotropy-Field used in the Oxsii 3-D solver in OOMMF was initialised with an Oxs_RandomVectorField object, causing each individual cell of the mesh to be magnetised initially with a random direction. When a magnetization run was continued over many computer sessions, the same anisotropy file was specified in every 'Specify Oxs_UniaxialAnisotropy' statement. This was found to reduce substantially the value of dm/dt (as defined in OOMMF) at the start of each session after the first, and so to speed up convergence slightly, in comparison with leaving the UniaxialAnisotropy-Field file unspecified. However, since the in-plane dimensions of a cell are probably smaller than the size of deposited grains, this procedure may have caused the simulations to produce more uniform results than the rings show experimentally. Use of a map with typical larger grains for the anisotropy file would be more realistic but might lead to a larger variation of computed results.

Simulations show clearly the different switching mechanisms for transitions from onion to vortex (O-V) and from vortex to onion (V-O) states. The O-V transition requires the movement of one Néel wall around half the circumference, while the V-O transition is started by the simultaneous rotation of all spins in a small region of the ring. For circular rings of the range of dimensions modelled, the O-V process was found to occur at a slightly smaller in-plane field than for V-O. Results obtained for the switching fields when a purely in-plane field is applied to cobalt rings of outer diameter 1 micrometre, with a thickness of 20 nm and a range of widths, are shown in Fig. 4. The figure shows that for the range of widths presented, there is a relatively narrow range of field that can switch the magnetisation into the vortex state. The effect of crystalline anisotropy described above may reduce this range further. With the anisotropy in the simulation limited as described above, there is more fluctuation in the O-V switching field than in the V-O field. Some non-circular shapes have been found to increase the usable range between switching fields, at the cost of requiring specific orientation in the magnetizing field, but they are not described here.

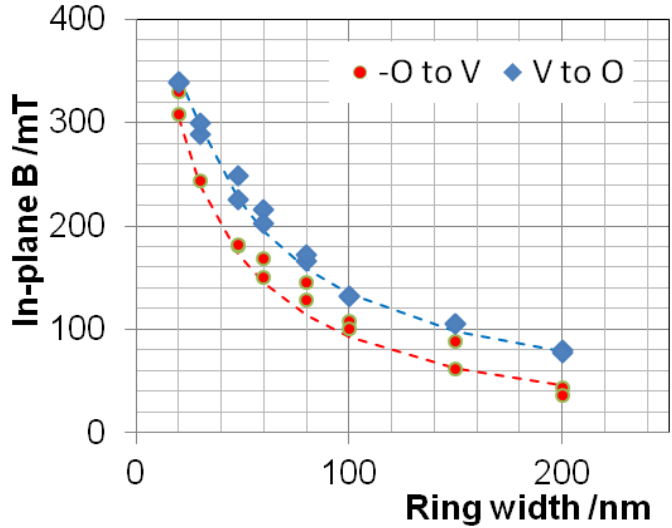

Fig. 4 Switching fields, onion to vortex (O-V) and vortex to onion (V-O), for cobalt rings of outside diameter 1 micrometre and thickness 20 nm, calculated using OOMMF [24], with approximate fit lines.

The time needed for a simulation increases rapidly with the ring diameter and with the number of cells in the axial (out-of-plane) direction. The magnitudes of in-plane switching fields are found to vary little with the overall diameter of the ring. For this reason simulations have been carried out for rings with an overall diameter close to 1 micrometre, and with one cell in the axial direction. The results have been used as a guide to the behaviour of rings with larger diameter.

The Data Table in the OOMMF software reports values of the Cartesian components of the magnetisation **M**. Thus values of $M_z$ can be determined easily, but the corresponding mean component of magnetization in the circumferential direction in the plane of the ring is not immediately available. It can be extracted (with some labour) for any cell from the detailed record of the magnetization for all cells. In the simulations reported here, the circumferential magnetization in the middle of the cross-section was found to agree closely with $(|M|^2 - M_z^2)^{1/2}$, as expected.

### 3.2 Fabrication

An ideal phase plate would consist of a ring that has no appreciable interception of scattered electrons, that does not require any supporting structure yet does not charge up in use, and that can be heated for cleaning if required. For the initial tests reported here, a simple structure was fabricated using the standard techniques of e-beam lithography and lift-off [26]. Resist was spun onto 50 nm low-stress silicon nitride (SiN) membranes supported on 2.9 mm silicon frames [27] and circular rings with diameters from 1 to 5.5 micrometres and nominal widths from 50 to 275 nanometres were written by e-beam lithography within the square window of 0.25 mm side. After development, the metallic layers were deposited by molecular beam epitaxy (MBE) in ultra-high vacuum (base pressure $3 \times 10^{-10}$ mbar) at room temperature, with a deposition rate below 2 monolayers per minute. This process was chosen because, in the ultra-high vacuum typical of MBE, the deposited atoms travel directly from the source to the substrate. If DC magnetron sputtering were used, the mean free path would be shorter than in MBE and so deposited atoms would arrive from a much larger range of directions. Then there would be more risk of bridging between the patterned rings and their unexposed surroundings, making it more difficult to remove the unwanted metallic areas and resist. We grew first an adhesion layer of 10 nm copper, then 20 nm of polycrystalline cobalt, and finally 3 nm of gold as an anti-oxidation/capping layer. All materials deposited had a purity level of 99.99%. The unwanted resist and metallic layer were removed by immersion in acetone. The calibration of the thickness of deposition of cobalt has been checked by x-ray reflectivity and found to be within 2% of the intended value of 20 nm.

For the first tests reported here, no anti-charging layer was used for the SiN. Consequently the beam current had to be kept lower than was desirable for good contrast of the holographic fringes. For later measurements an anti-charging coating of carbon was added by arc deposition.

## 4. Results of holography

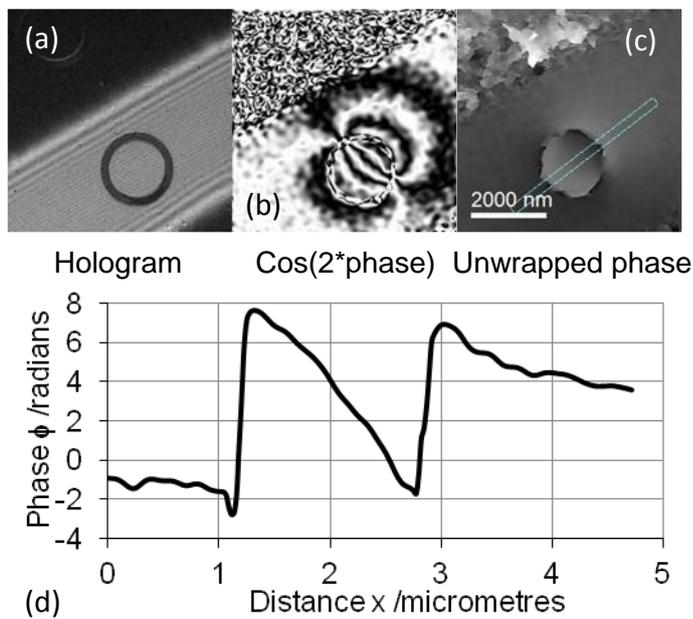

Fig. 5 Results of holographic measurement on a cobalt ring of outside diameter 1.9 micrometres, width 217 ± 6 nm and thickness 20 nm: (a) hologram; (b) cosine of 2 x phase (integrated along the beam path); (c) unwrapped phase with outline of a line scan; (d) plot of integrated phase along the line scan.

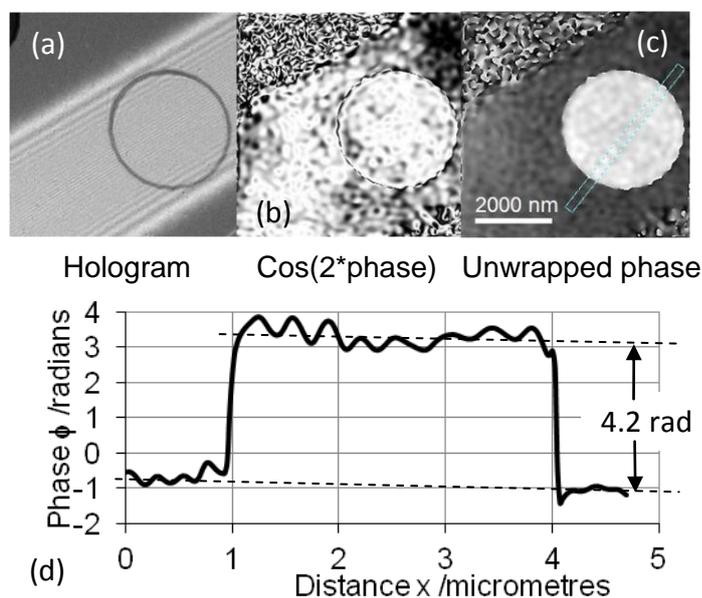

Fig. 6 Results of holographic measurement on a cobalt ring of outside diameter 3.1 micrometres, width 83 ± 7 nm and thickness 20 nm: (a) hologram; (b) cosine of 2 x integrated phase; (c) unwrapped phase showing a line scan; (d) plot of integrated phase along the line scan.

A frame containing rings of the diameters and widths listed above was magnetized before insertion in the transmission electron microscope by applying 400 mT in one in-plane direction, then 124 mT in the opposite direction. The second field was chosen with the aim of setting up the vortex state in a ring with a width of 100 nm. Measurements were made with a CM300 microscope equipped for holography. The objective lens was not excited for the tests reported here, so images were obtained by use of the diffraction lens (part of the intermediate assembly). The field at the specimen was less than 2 mT. In Figures 5(a), 6(a) and 7(a) the biprism wire runs across each image from lower left to upper right and can be identified by its characteristic Fresnel fringes.

Figure 5 shows results for a ring whose outside diameter and width are measured as 1.9 micrometres and 217 nanometres ± 6 nm. In Fig. 5(b), the most obvious fringes provide contours of phase showing the direction of $B_t$, the integrated component of **B** transverse to the beam, as described in the Appendix. Fig. 5(c) shows, superimposed on the image, the path of a scan which recorded the phase $\theta$ as a function of distance x, as shown in Fig. 5(d). The rate of change of phase with scan distance, $d\theta/dx$, is proportional to the component of $B_t$ *perpendicular* to the direction of the scan. It can be seen that there are two positions, separated by the diameter of the ring, at which $d\theta/dx$ is large and positive. Since $d\theta/dx$ has the same sign at both these positions, $B_t$ has the same (Cartesian) direction on both sides of the ring, showing that the ring was magnetised in the onion state. Between these positions, $d\theta/dx$ is negative and smaller in magnitude, showing that in the space within the ring, **B**$_t$ is in the opposite direction to that in the material of the ring. This is expected for an onion state; the ring has a magnetic moment of dipole form, and flux returns outside the material of the ring in the opposite direction to that in the ring.

The behaviour of a ring with outside diameter and width measured as 3.1 micrometres and 83 ± 7 nanometres is shown in Fig. 6. In contrast to the previous figure, the image and graph of unwrapped phase (Fig. 6(c) and (d)) show that the phase is approximately constant across the whole interior of the ring, with no divergence of flux. The phase difference between inside and outside of the ring is measured as 4.2 ± 0.5 radians. It appears from Fig. 6(d) that this phase difference is similar around the entire circumference, which implies that the flux in the ring has the same angular direction everywhere and the ring is in the vortex state.

Figures 5 and 6 show that the two rings were switched into the two different states by the same magnetising field, and therefore confirm that at least one switching field varies with the ring width.

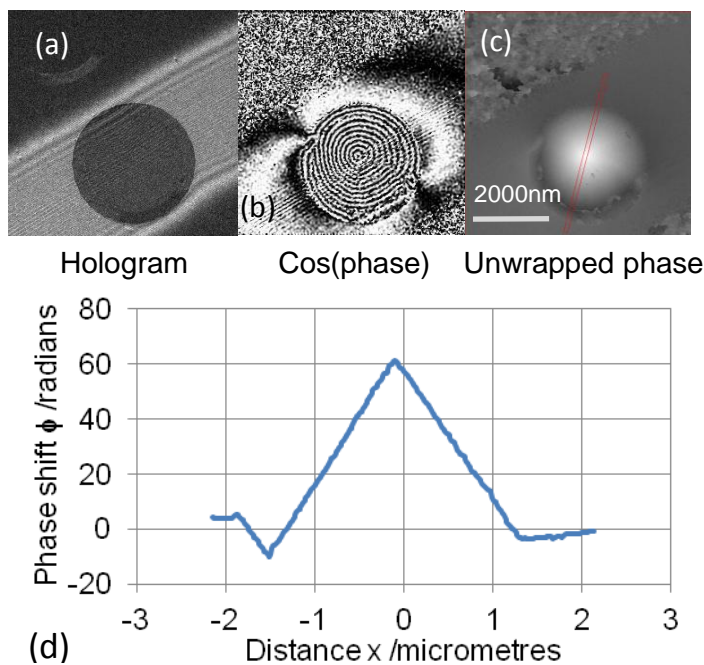

Fig. 7  Holographic measurements for a cobalt ring of outside diameter 3.2 micrometres and thickness 20 nm, together with its central disc displaced by a few nanometres from the centre of the ring: (a) hologram; (b) cosine of integrated phase; (c) unwrapped phase showing a line scan; (d) plot of integrated phase along the line scan.

The behaviour of a combination of a ring together with its central disc was also recorded (Figs. 7 and 8). In this case the central disc was not removed by the lift-off process but remained close to its as-deposited position. This geometry is not usable for a phase plate, but the results are valuable because they provide quantitative measurement of the flux density in the ring and show that this is near the bulk saturation value. The cos(phase) plot (Fig. 7(b)) shows a series of concentric near-circular loops, with an additional arc at the lower left. The scan of phase against distance shown in Fig. 7(c) passes through the centre of this pattern of circles. The corresponding plot of the phase (Fig. 7(d)) shows two broad regions of similar spatial extents and opposite slopes. The similar total extents of phase for both signs of slope show that the totals of flux circulating on opposite sides of the centre are similar in magnitude but opposite in direction, and thus form a vortex state. The additional arc at the lower left is magnetised in the opposite direction to the part of the vortex that is nearest to it. This is clearly shown by the slope of Fig. 7(d) at scale distances between −1.8 and −1.52 micrometres and in Fig. 8(a) by the colour coding that shows the direction of magnetization. The detailed phase measurements show that the total phase change for the two regions of negative slope in Fig. 7(d) exceeds that for the region of positive slope by about 12%. The spatial distribution of phase ranges (proportional to flux) is indicated in Fig. 8(b). Thus the total magnetic flux above and to the right of the centre of the circular pattern in Fig. 7(b), together with the flux in the 'extra' arc at lower left, also exceeds the flux in the opposite direction, below and to the left of the centre of the pattern of circles. The excess flux returns outside the disc and ring, as

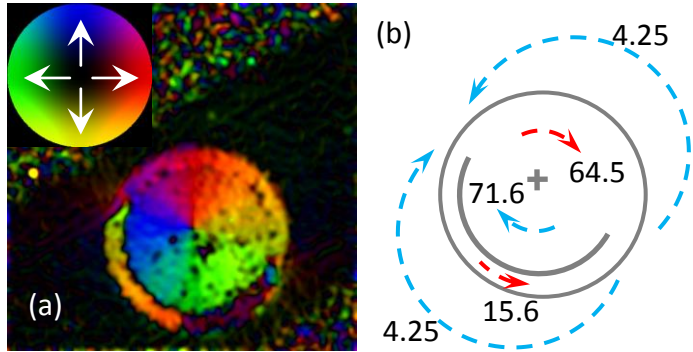

Fig. 8 (a) The ring as in Fig. 7 with the in-plane direction of flux (integrated along electron paths) mapped to colour; with colour wheel (inset): a particular colour shows that the direction of flux is the direction of a line from the centre of the wheel to that colour; (b) distribution of flux directions and phase extents deduced from the phase scan shown in Fig. 7. The numbers show phase extents in radians, proportional to flux, for the sections with constant slope in the plot of Fig. 7(d), and for the external flux needed for continuity.

can be seen from the dipole pattern outside the ring system in Fig. 7(b). This external flux is about 45% of that in the arc at lower left. The flux densities within the disc, calculated from the phase gradients, are slightly non-uniform; with the assumption that flux is confined to the thickness of cobalt, the average flux density in the lower part of the disc is 1.66 T, and in the upper part is 1.52 T. The larger of these flux densities is within 8% of the value of 1.79 T known for the saturation flux density of cobalt in bulk [23].

Under enlargement, the original of Fig. 7(a) shows a gap of the order of 30 nm between the ring and the disc at the lower left side, and a corresponding overlap at the upper right. It is also possible that some residue of the resist remained to raise the disc slightly from the ring. A separation greater than 5 nanometres would greatly reduce the exchange coupling of the spins in ring and disc, and so would permit the opposition of flux directions on the two sides of the gap.

## 5. Discussion

Holographic imaging of narrow rings has shown clearly that use of a single magnetizing field can produce onion or vortex states, depending on the ring width as well as field, in agreement with previous work [21].

The phase difference produced by the ring in the vortex state (Fig. 6), of width 83 ± 7 nm and thickness 20 ± 0.4 nm, was measured as 4.2 ± 0.5 radians. The phase shift calculated for these dimensions, using the mean of flux densities deduced from measurements on the disc-and-ring system, is (flux density) x (area) x $e / \hbar$ = 4.0 ± 0.4 radians. Thus the experimental phase difference produced by the ring in the vortex state (Fig. 6) agrees within measurement error with the phase expected theoretically from these dimensions.

The range of magnetizing field that can be used to switch a ring of given width into the vortex state is small, and may be limited further by the magneto-crystalline anisotropy. Other shapes of ring that have been investigated may have greater ranges of switching field, but they are not discussed here. Also, the results shown here are for a magnetizing field applied entirely in-plane. The response of rings to additional out-of-plane fields requires further study. The simulations carried out for the present work, using a single cell in the axial direction, suggest that the vortex state will withstand fields of the order of 1 T , as proposed earlier [17]. However, they also suggest that a specimen tilt relative to this field by an angle of the order of 5 degrees will reduce the V-O switching field appreciably. On the other hand, detailed investigations with rings of cobalt nanoparticles have shown [28] that they can form a 'double vortex' structure, stacked in the axial (out-of-plane) direction, and that the vortex state can be switched between stable states of opposite chirality by the application of axial fields of the order of 2 T. It is clearly desirable to carry out similarly detailed studies to find out how the properties of thin-film rings of varying thicknesses are related to those of rings of discrete particles.

In the ring and disc system of Fig. 7, much of the disc presents a clear vortex pattern but the detailed phase measurements show that one cannot say simply that the outer ring is in the onion state, as was suggested earlier [7]. About half the flux in the extra arc returns by joining that nearby in the major vortex, and the rest returns externally.

The results presented here have been obtained from structures that are relatively simple to fabricate. It is clear that for use of a similar structure in a microscope column, further development is needed to limit the effect of charging, possibly including provision for heating. Also for regular use, a more robust support is needed than the silicon nitride membrane used here. The major benefits of such membranes for initial trials are of course their ready availability [27] and their near-transparency to electrons, which eases the imaging of phase by holography.

## 6. Conclusions

Holographic imaging has demonstrated clearly the onion and vortex states of magnetised ferromagnetic thin-film rings. The distributions of phase of the electron wave function produced by these states can be distinguished easily by holography. A ring in the vortex state produces a uniform phase difference between paths through the interior and exterior of the ring, as expected from Aharonov-Bohm theory. The magnitude of this phase difference agrees very well with the value predicted by theory, within measurement error. The radial width of a cobalt ring needed for $\pi/2$ phase difference is just under 30 nm, suggesting that when a ring of this type is used as a phase plate, interception of scattered electrons can be limited to a similarly small radial range.

## 7. Appendix

According to Aharonov-Bohm theory as in Section 2 above, two trajectories with the same values of $\int_p^q \mathbf{A} \cdot d\mathbf{L}$ have the same values of phase shift due to $\mathbf{A}$, and the net flux through a surface bounded by them is zero. A sequence of such trajectories defines a surface of constant phase. The intersection of this surface with a plane cross-section of the system defines a contour line of constant phase. There is no net flux through the surface containing this contour, and so the direction of flux density must be parallel to this surface. So the direction of $\mathbf{B_t}$, the (integrated) transverse component of flux density $\mathbf{B}$, is the same as the local direction of the contour in the plane of the cross-section, as seen in Fig. 5(b). Between contours of equal increments in phase there are equal increments in flux, so the magnitude $|\mathbf{B_t}|$ is inversely proportional to the spacing of phase contours. With the assumption that the transverse flux is confined to the thickness t of ferromagnetic material, $\mathbf{B_t}$ can be obtained by use of equation (1) as

$$\mathbf{B_t} = (\hbar/e\,t)(d\theta/dx)\hat{\mathbf{S}}$$

where $\hat{\mathbf{S}}$ denotes a unit vector in the transverse direction of the phase contour, and dx is measured transversely and perpendicular to $\hat{\mathbf{S}}$. Thus in Fig. 5(b), the broad spacing of contours inside and outside the ring has a dramatic appearance but actually shows that the flux density there is smaller than within the material of the ring, where the contours are more closely spaced.

The vector potential $\mathbf{A}$ for this static system is a solution of Poisson's equation in which the source term is proportional to the magnetization current density, $\mathbf{J}$, that results from the summation of currents on the atomic scale. Consider a ferromagnetic thin-film ring and a test plane through its axis. The plane intersects the ring in two small rectangles. When the ring is magnetised in either the onion or the vortex state, as shown in Fig. A.1, the atomic magnetic moments in these rectangles are aligned perpendicular to the test plane. They correspond to small loops of current whose adjacent parts cancel over most of the rectangular section. The non-cancelled parts of these small loops of current form loops, mainly around the perimeters of the rectangles. These loops of current provide the source that generates **A** according to Poisson's equation. It is sometimes stated that the distribution of **A** is 'like **J**, but fuzzier', and this suggests that (with the choice of the Coulomb gauge, div A = 0) there are loops of **A** around the magnetic material at all azimuthal angles, the sense of circulation around the material corresponding to the direction of magnetisation of the appropriate half of the ring. Thus close to the ring, the loops of **A** combine to form toroidal surfaces. However, at points further from the ring, there are contributions from all elements of the ring and to find the distribution of **A** qualitatively it is simpler to consider the experimental flux distribution.

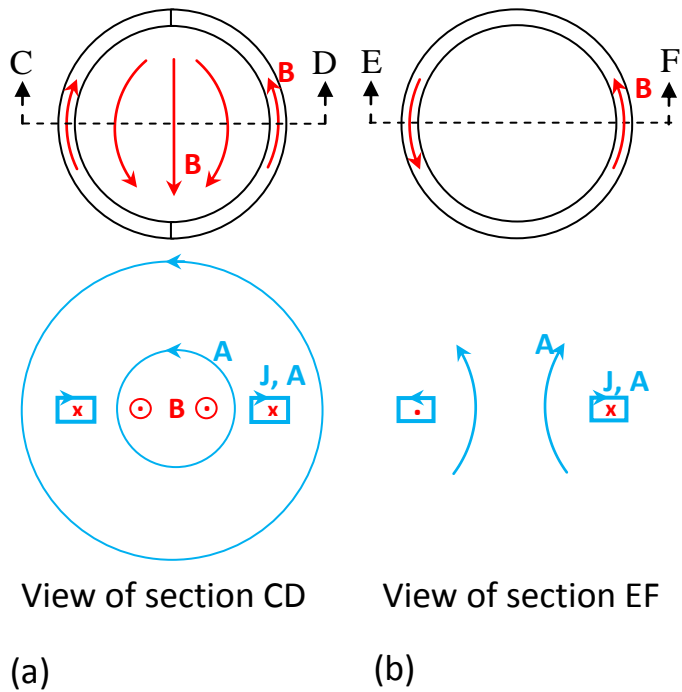

Fig. A.1 Distribution of flux density **B** and vector potential **A** for magnetization current **J** in a thin narrow ring (a) in the onion state and (b) in the vortex state. The directions of **B** normal to the plane are shown by 'x' (into the plane) or '.' (out of the plane).

The ring shown in Fig. 5 can be seen to have a flux distribution like that of a dipole, as expected for the onion state. Fig. 5(d) shows that flux returns through the centre of the ring in the opposite (Cartesian) direction to that in the magnetic material. For consistency with the relation **B** = curl **A**, and again with the assumption for this static system that div **A** = 0, the central flux must be surrounded by loops of **A,** but the direction of circulation for these loops must be opposite to those in the ferromagnetic material. Thus, for the onion state, the distribution of **A** in the equatorial plane is of the form shown in Fig. A.1(a).

For the vortex state, where the system has rotational symmetry, **J** has no azimuthal (or tangential) component and so neither does **A**. Loops of **A** surround the magnetised material but their magnitude does not vary azimuthally. Ampere's law shows that at any point that is not in the magnetised ring, the tangential component of **B** is zero. It can then be shown by symmetry that the other components of **B** are also zero outside the material of the ring.

## 8. Acknowledgments

The authors thank Justin Llandro and Theodossis Trypiniotis for magnetisation of the specimens used. The work was organised within the Thin-Film Magnetism and Materials Group of the Department of Physics of Cambridge University.

**Contributors**

**Funding**
JCL holds a University Research Fellowship from the Royal Society. The lithography by AMB was funded by Hitachi Europe Ltd. Experimental fabrication was supported by the Thin Film Magnetism group of C.U. Department of Physics. Those authors who are not members of Hitachi Lab have no directly relevant financial links with any commercial or scientific organization or other funding body.